\documentclass[twocolumn,superscriptaddress,prd]{revtex4-1}
\usepackage{bm}
\usepackage{color}
\usepackage{amsmath}
\usepackage{natbib}
\usepackage{graphicx}
\usepackage{cancel}
\usepackage{longtable}
\definecolor{RED}{rgb}{1,0,0}
\begin{document}
\title{Radiative spin polarization of electrons in an ultrastrong magnetic field}
\author{Koen van Kruining}
\email{koen@pks.mpg.de}
\author{Felix Mackenroth}
\email{mafelix@pks.mpg.de}
\affiliation{Max Planck Institut f\"ur Physik komplexer Systeme, 01187 Dresden, Germany}
\author{J\"org B. G\"otte}
\affiliation{University of Glasgow, Glasgow, G12 8QQ, UK}
\affiliation{Nanjing University, 210093 Nanjing, China.}
\begin{abstract}
We calculate the spin flip rates for an electron in a homogeneous magnetic field for low excitations ($N\le 5$). Our results apply for all field strengths including those beyond the critical field strength at which the spin contributes as much to the electron's energy as its rest mass. Existing approximations either assume that the electron is in a sufficiently highly excited state such that its orbit can be assumed to be classical or the magnetic field be weak compared to the critical field. The regime of high magnetic field strength and low excitations is therefore poorly covered by them. By comparing our calculations to different approximations, we find that in the high field, low excitation regime the spin flip rates are lower and the equilibrium spin polarization is less pure then one would get by naively applying existing approximations in this regime.
\end{abstract}
\maketitle
\section{Introduction}
A relativistic electron in a homogeneous magnetic fields is one of the oldest solved problems in relativistic quantum mechanics. Mere months after Dirac presented his equation, Rabi found an incomplete solution allowing only for one spin orientation relative to the magnetic field \cite{Rabi1928}. Landau studied the problem extensively to explain diamagnetism and introduced a set of energy levels associated with the total angular momentum of the electron \cite{LL3, LL4} and Sokolov \& Ternov (S\&T) used a complete set of relativistic solutions extensively to study quantum effects in synchrotron radiation emission \cite{ST}. Recently the similarity to free space electron vortex beams \cite{BliokhBliokhSNori07, *BliokhDennisNori11, vanBoxemVerbeeckPartoens13, Barnett17} led to a renewed interest in Landau states \cite{BliokhSchattschneiderVNori12, GreenshieldsStampsFranke-Arnold12, *GreenshieldsFranke-ArnoldStamps15, GSchattshneiderBliokhNVerbeeck13, *SchattschneiderSS-PLofflerS-TBliokhNori14, vanKruiningHayrapetyanGotte17, SilenkoZhangZou18}

The S\&T calculations for the spin flip rates in a magnetic field predict that the equilibrium spin polarization of an electron in a highly excited quantum state is $(n_\downarrow-n_\uparrow)/(n_\downarrow+n_\uparrow)=8\sqrt3/15\approx0.924$ \cite{ST, LL4}. S\&T approximated the Laguerre-Gau\ss{} wave functions of the electrons by Bessel functions, an approximation which breaks down at low Landau levels where the finite widths of the electron wave functions become important.

Ternov, Bagrov \& Dorofejev (TB\&D) computed the spin flip rates for low excitations assuming a magnetic field which is weak compared to the critical field strength $B_\text{cr}=m_e^2/|e|\approx 4.4\mbox{ GT}$ which yields an equilibrium spin polarization of 
\begin{equation}
\frac{n_\downarrow-n_\uparrow}{n_\downarrow+n_\uparrow}=\frac{(m_e+2B|e|N)^2-4a B^2|e|^2N^2}{(m_e+2B|e|N)^2+4a B^2|e|^2N^2}
\end{equation}
with $B$ the magnetic field strength, $m_e$ the electron mass, $N$ the principal quantum number of the state in question and $e$ the electron's charge. The dimensionless parameter $a$ is introduced by TB\&D with the explanation `where the numerical factor $a$ does not exceed unity'  \cite{TernovBagrovDorofejev68}. Because they give no further indication about the magnitude of $a$, we will treat it as a free parameter to be fitted.

For typical experiments in a Penning trap \cite{HannekeHoogerheideGabrielse11} the magnetic field strength does not exceed 10 T and the TB\&D-approximation works well. In storage rings, which use similar magnetic fields but much higher kinetic energies for the electrons, one can assume that the electrons can transition into a continuum of states, instead of a discrete spectrum, yielding the S\&T-approximation\cite{ST,LL4}. New proposals for generating short-lived magnetic fields in plasmas created by ultrashort laser pulses allow for field strengths of 0.1-1 MT\cite{LeczKSerjiAndrejev16, StarkToncianArefiev16}, which is three to four orders of magnitude beyond the field strengths that are available from non-destructive magnets \cite{Simsetal08}. Even stronger magnetic fields, up to $10^{11}$ T can be found on the surface of neutron stars \cite{Paczynski92, ThompsonDuncan95, *ThompsonDuncan96, VasishtGotthelf97, Kouveliotouetal98}. 

For such strong fields, the spacing between Landau levels is larger than the electron's rest mass and exciting an electron to a high enough Landau level such that the S\&T-limit is reached requires an excessive amount of energy. The TB\&D approximation explicitly assumes a magnetic field weak compared to the critical field and does therefore not apply at these field strengths either \cite{TernovBagrovDorofejev68}. Thus full nonperturbative QED techniques are required \cite{Ritus85, Baier_b_1998}, making no additional assumptions. A suitable non-perturbative QED theory can be described in the Furry picture \cite{Furry51} through replacing the vacuum electron states by solutions of the Dirac equation in a magnetic field \cite{ST, MelroseParle83a, vanKruiningHayrapetyanGotte17}. The Furry picture has been applied to the electron in a magnetic field to study vacuum birefringence \cite{Minguzzi56, Constantinescu72a}, energy corrections to the Landau levels \cite{Constantinescu72b, Parle87, GRHeroldRuderWunner94}, Compton scattering \cite{Herold79} and synchrotron radiation emission \cite{HeroldRuderWunner82, Latal86, HardingPreece87, *BaringGonthierHarding05, BezchastnovPavlov91, *PavlovBezchastnovMA91, SemionovaLeahyPaez10}. Analogous studies of radiation emission in this nonlinear QED regimes were previously conducted in the interaction of relativistic electrons with ultra-intense laser fields \cite{MackenrothDiPiazza11, *DiPiazza_etal_2012, *MackenrothNeitzDiPiazza13, *AngioiMackenrothDiPiazza16}, a regime which is complementary to the pure magnetic field case studied here. In the absence of strong fields, the scattering theory of non-plane wave states has been explored \cite{Ivanov11, * IvanovSerbo11, *Ivanov12, *Serboetal15, *Matulaetal14, *IvanovSSurzhykovFritzsche16, *ZaytsevSerboShabajev17, SchattschneiderLS-PVerbeeck14, *JuchtmansBABVerbeeck15, *JuchtmansVerbeeck15,BogdanovKazinskiLazarenko18, *BogdanovKazinskiLazarenko19} using techniques which are useful in strong-field situations too.

In this article, we analytically investigate and numerically compute the spin flip rates of low lying Landau levels and compare and contrast these results with both the S\&T- and TB\&D-approximations.

\section{Analytical model}
We will take the magnetic field to be constant, homogeneous with field strength $B$ and pointing in the positive z-direction, which is the quantization axis for angular momenta too. We use $\eta_{\mu\nu}=\mbox{diag}(+---)$ as our metric, express energies in eV and use $c=\hbar=1$ and $\epsilon_0=\mu_0^{-1}= 1/4\pi$, which implies $|e|=\sqrt\alpha\approx 1/\sqrt{137}$. 

For this geometry there exists a complete exact basis of nondiffracting Laguerre-Gau\ss{} beam solutions with their beam axes along the magnetic field \cite{ST, MelroseParle83a,vanKruiningHayrapetyanGotte17}. We chose the spin states to be eigenstates of the component of the magnetic moment operator $\vec \mu=m_e\vec\Sigma-i\vec\gamma\times (-i\vec\partial-e\mathbf A)$ along the beam axis  \cite{ST, HeroldRuderWunner82,MelroseParle83a}. Here $\vec\Sigma$ is a vector of $4\times4$ spin matrices and $\vec\gamma$ a vector of gamma matrices. We define the Laguerre-Gau\ss{} functions $LG_n^l(r,\phi)=e^{il\phi}r^{|l|}e^{-\frac{r^2}2}L_n^{|l|}(r^2)$, $L_n^l$ is an associated Laguerre polynomial and $\tilde r=\sqrt{B e/2}r$ to write the electron states as
\begin{widetext}\begin{equation}
\Psi=\frac{e^{i(pz-\mathcal E t)}}{\sqrt{\mathcal N}}
\begin{cases}
\left[\begin{array}{c}
(\mathcal E+\mathcal E_0)\sqrt{\mathcal E_0+m}\,LG_n^l(\tilde r,\phi)\\
-ip\sqrt{\mathcal E_0-m}\,LG_n^{l+1}(\tilde r,\phi)\\
p \sqrt{\mathcal E_0+m}\,LG_n^l(\tilde r,\phi)\\
 i(\mathcal E+\mathcal E_0)\sqrt{\mathcal E_0-m}\,LG_n^{l+1}(\tilde r,\phi)
\end{array}\right]& l\ge0,\sigma>0\\
\left[\begin{array}{c}
-ip \sqrt{\mathcal E_0-m}\,LG_n^{l-1}(\tilde r,\phi)\\
(\mathcal E+\mathcal E_0) \sqrt{\mathcal E_0+m}\,LG_n^l(\tilde r,\phi)\\
-i(\mathcal E+\mathcal E_0) \sqrt{\mathcal E_0-m}\,LG_n^{l-1}(\tilde r,\phi)\\
-p\sqrt{\mathcal E_0+m}\,LG_n^l(\tilde r,\phi)
\end{array}\right] &
l>0,\sigma<0\\
\left[\begin{array}{c}
(\mathcal E+\mathcal E_0)\sqrt{\mathcal E_0+m}\,LG_n^l(\tilde r,\phi)\\
ip \sqrt{\mathcal E_0-m}\,LG_{n+1}^{l+1}(\tilde r,\phi)\\
p\sqrt{\mathcal E_0+m}\,LG_n^l(\tilde r,\phi)\\
-i(\mathcal E+\mathcal E_0)\sqrt{\mathcal E_0-m}\,LG_{n+1}^{l+1}(\tilde r,\phi)
\end{array}\right]& l< 0,\sigma>0\\
\left[\begin{array}{c}
ip \sqrt{\mathcal E_0-m}\,LG_{n-1}^{l-1}(\tilde r,\phi)\\
(\mathcal E+\mathcal E_0)\sqrt{\mathcal E_0+m}\,LG_n^l(\tilde r,\phi)\\
i(\mathcal E+\mathcal E_0)\sqrt{\mathcal E_0-m}\,LG_{n-1}^{l-1}(\tilde r,\phi)\\
-p\sqrt{\mathcal E_0+m}\,LG_n^l(\tilde r,\phi)
\end{array}\right]& l\le0,\sigma<0.
\end{cases}
\end{equation}\end{widetext}
They are specified by a momentum along the beam axis, $p$, radial quantum number, $n$, orbital quantum number, $l$, and spin $\sigma=\pm\frac 12$. Although spin-orbit mixing makes it impossible to attribute an integer orbital angular momentum and a half-integer spin to a given state, the labeling with integer $l$ and half integer $\sigma$ reflects the limiting behavior $B\rightarrow 0$. The total angular momentum $j=l+\sigma$ is always half-integer. The energy of the electron is $\mathcal E=\sqrt{m_e^2+p^2+2 B|e|N}$ with $N= n+\frac 12(l+|l|)+\sigma+\frac 12$. The electron states are normalizable in the transverse plane with normalization $\mathcal N=4\pi\times 4\mathcal E\mathcal E_0(\mathcal E+\mathcal E_0)/|e|B\times (n+|l|)!/n!$. Here $\mathcal E_0=\sqrt{m_e^2+2 B|e|N}$.

The most effective way to keep track of angular momentum changes of the electron when it radiates is to expand the photon field too in a basis of eigenstates of angular momentum along the beam axis. These are the photon Bessel modes with total angular momentum $j_\gamma$, momentum along the beam axis $k$, transverse momentum $\kappa$ and energy $\omega=\sqrt{k^2+\kappa^2}$ . For the photon polarization we use a basis of left- ($-$) and right- ($+$) handed helicity. For a photon emitted in the positive z-direction positive helicity corresponds to a predominantly positive spin whereas for a photon emitted backward it corresponds to a predominantly negative spin, with the expectation value for the photon spin along the beam axis continuously decreasing with decreasing $k$. Using the Coulomb gauge these photon modes are described by the vector potential
\begin{widetext}\begin{equation}
\mathbf A_{k\kappa j\pm}=\frac {e^{i(kz-\omega t)}}{2\sqrt \mathcal N_\gamma}\left[\begin{array}{c}
\left(1\pm \frac k\omega\right)J_{j_\gamma-1}(\kappa r)e^{i(j_\gamma-1)\phi}+\left(1\mp\frac k\omega\right)J_{j_\gamma+1}(\kappa r)e^{i(j_\gamma+1)\phi}\\
i\left(\left(1\pm\frac k\omega\right)J_{j_\gamma-1}(\kappa r)e^{i(j_\gamma-1)\phi}-\left(1\mp\frac k\omega\right)J_{j_\gamma+1}(\kappa r)e^{i(j_\gamma+1)\phi}\right)\\
\mp 2i \frac \kappa\omega J_{j_\gamma}(\kappa r)e^{ij_\gamma\phi}
\end{array}\right].
\end{equation}\end{widetext}
With these states, one can compute the transition matrix element $\mathcal M=\int\bar\Psi_f\cancel A^*\Psi_idV$ with $\bar\Psi_f=\Psi_f^\dagger\gamma_0$, and slash denoting contraction with the Dirac matrices $\cancel A^*=\gamma^\mu A_\mu^*$. The explicit forms of $\mathcal M$ are given in the appendix. Performing the integrations over the coordinate along the beam axis, time and the azimuthal angle will yield three delta functions which we will take out of $\mathcal M$. Then, only the radial integration remains. Unlike for plane waves, there is no fourth delta function as there is no fourth conserved quantity whose operator commutes with the angular momentum operator and therefore the electron and photon states cannot be simultaneous eigenstates of four conserved quantities. To compute the decay rate from the state $N,j,\sigma$ to $N',j',\sigma'$ with $N>N'$, we have to compute the squared transition matrix element and integrate it over all outgoing coaxial electron momenta, all coaxial and transverse photon momenta and sum over the photon's angular momentum and both photon polarizations. We take the initial coaxial momentum of the electron $p=0$ for definiteness. The decay rates for an electron moving along the beam axis can be found by a Lorentz transformation.  Taking our wave functions confined to a disc of thickness $L$ and radius $R$ (which we will take to infinity), the outgoing electron's density of states is $L/2\pi$. The photon's density of states (for a single angular momentum and polarization) is $L/2\pi\times R/\pi$. Putting all these ingredients together, the decay rate is
\begin{widetext}\begin{equation}\label{eq:diffdecrate}
\Gamma_{Nj\sigma\rightarrow N'j'\sigma'}=(2\pi)^4e^2\sum_\pm\sum_{j_\gamma}\iiint\delta(\mathcal E-\mathcal E'-\omega)\delta(p-p'-k)\delta_{j,j_\gamma+j'}L\frac{|\mathcal M|^2}{L\mathcal N L\mathcal N'L\mathcal N_\gamma}\frac{Ldp'}{2\pi} \frac{LRdk d\kappa}{2\pi^2}.
\end{equation}
Primes refer to the properties of the final electron state. Using the asymptotic form for $r\gg\kappa^{-1}$ of the Bessel functions $J_j(\kappa r)\approx\sqrt{2/\pi\kappa r}\cos(\kappa r-(j+\frac 12)\frac\pi 2)$ and the normalization condition $\int\mathbf A\cdot\mathbf A^*dV=2\pi/\omega$, $\mathcal N_\gamma$ can be computed for $R\gg \kappa^{-1}$. We find $\mathcal N_\gamma\approx R\omega/\pi\kappa$.  Substituting the transverse normalization factors in eq.~(\ref{eq:diffdecrate}), $L$ and $R$ disappear and the size of the disc can be taken to infinity, yielding
\begin{equation}
\Gamma_{Nj\sigma\rightarrow N'j'\sigma'}=(2\pi)^4e^2\sum_\pm\sum_{j_\gamma}\iiint\delta(\mathcal E-\mathcal E'-\omega)\delta(p-p'-k)\delta_{j,j_\gamma+j'}\frac{|\mathcal M|^2}{\mathcal N \mathcal N'}\frac{dp'}{2\pi} \frac{\kappa dk d\kappa}{2\pi}.
\end{equation}\end{widetext}
The Kronecker delta of the angular momenta will be eliminated by summing over all photon angular momenta. The delta function of the coaxial electron momentum will be eliminated by integrating. To treat the delta function of the energies, we rewrite the photon momentum in polar coordinates, $\kappa=\omega\sin\theta$, $k=\omega \cos\theta$ and $d\kappa dk=\omega d\theta d\omega$. Using energy and momentum conservation, we have $\mathcal E'=\sqrt{m_e+2B|e|N'+\omega^2\cos^2\theta}$. Now integrating over $\omega$ and eliminating the energy delta function gives an additional factor of
\begin{equation}
\frac 1{\left|\frac{d \mathcal E-\mathcal E'-\omega}{d\omega}\right|}=\frac{\mathcal E'}{\mathcal E'+\omega\cos^2\theta}=\frac{\mathcal E-\omega}{\mathcal E-\omega\sin^2\theta}.
\end{equation}
the effect of this factor has been pointed out before \cite{ST, HeroldRuderWunner82} and it will be especially important in the strong field regime where we expect the S\&T- and TB\&D-approximations to break down. Including this factor, the phase space integral is
\begin{equation}
\Gamma_{Nj\sigma\rightarrow N'j'\sigma'}=(2\pi)^2e^2\sum_\pm\int_0^\pi\frac{|\mathcal M|^2}{\mathcal N \mathcal N'} \frac{(\mathcal E-\omega)\omega\sin\theta d\theta}{\mathcal E-\omega\sin^2\theta}.
\end{equation}
The remaining integral has to be integrated numerically.

\section{Numerical results}
We numerically computed all transition rates for electron states in a magnetic field with $N\le 5$ and $j\ge -9/2$ to final states with $N'\le 4$ and $j'\ge-17/2$. The highest allowable $j$ for a given $N$ is always $j=N-1/2$. The reason to choose a lower $j$ cutoff for the final state is that the electron tends to lose angular momentum when it decays more often than it gains it. To compute the spin flip rate we take the sum $\Gamma_{\sigma\sigma'}=\sum_{N'j'}\Gamma_{Nj\sigma\rightarrow N'j'\sigma'}$ over all energetically allowed final states in this data set. This yields two spin flip rates for a given initial $N$ and $j$ from which we compute an approximation for the equilibrium spin polarization according to the model in Appendix~\ref{app:teq}. As the electrons radiate, a constant input of energy is needed to keep them on the same energy level. We make no assumption about the mechanism by which the electrons are reenergized, except that it does not cause spin flips, as is customary \cite{ST, TernovBagrovDorofejev68}. Because we obtain an equilibrium spin polarization fairly close to one, we plot for readability $2n_\uparrow/(n_\uparrow+n_\downarrow)$, which is the deviation of the equilibrium spin polarization from unity.

To compare our results for the relative spin flip rates of electrons to the S\&T results, we use the plane wave approximation from \cite{LL4}. Here the electron is considered to be a plane wave with the time evolution $e^{-iHt}$, with $H$ the Hamiltonian of the problem under consideration. This approximation takes the recoil of the electron when it emits a photon into account properly and is thus an improvement over classical electrodynamics if $\mathcal E\gtrsim m_e^3/|e|B$. It reproduces the S\&T spin flip rates exactly \cite{LL4}. We choose a plane wave momentum equal to the expected tangential momentum of the highest $l$ state for a given $N$. This momentum can be written using Euler gamma functions.

\begin{equation}
p_\perp=\sqrt{2B|e|}\frac{\Gamma(N+\frac 12)}{\Gamma(N)}.
\end{equation}
Interestingly, taking this expression for the tangential momentum of the electron implies that in the weak field limit the change of momentum the electron gets upon decaying far exceeds the emitted photon's momentum, which is $\sim B|e|/2m_e$. Thus upon emitting a photon, the electron always gets a `superkick' \cite{BerryBarnett13}, caused by the spatial extent of the photon mode being much larger than the spatial extent of the electron's wave function. When angular momentum is transferred from one to the other, the electron's tangential momentum must change by much more than the emitted photon's tangential momentum. Because the electron's tangential momentum averaged over one orbit is zero, momentum is nonetheless conserved.

For comparison with the TB\&D results we took their expressions for the spin flip rates \cite{TernovBagrovDorofejev68}
\begin{equation}
\Gamma_{flip}=\frac 52\frac{m_e ce^2}{\hbar^2 4\pi\epsilon_0}\left(\frac B{B_\text{cr}}\right)^3
\begin{cases}
1 & \sigma<0\\
\frac{4a B^2|e|^2N^2}{(m_e+2B|e|N)^2} & \sigma>0
\end{cases}
\end{equation}
and fit them to our numerical spin flip rates for field strengths up to $10^7$ T to find the numerical value for $a$. Because in the TB\&D approximation, one can replace $\tilde B=NB$, one can fit the data of different Landau levels to the same curve.
\begin{figure}
\includegraphics[width=\columnwidth]{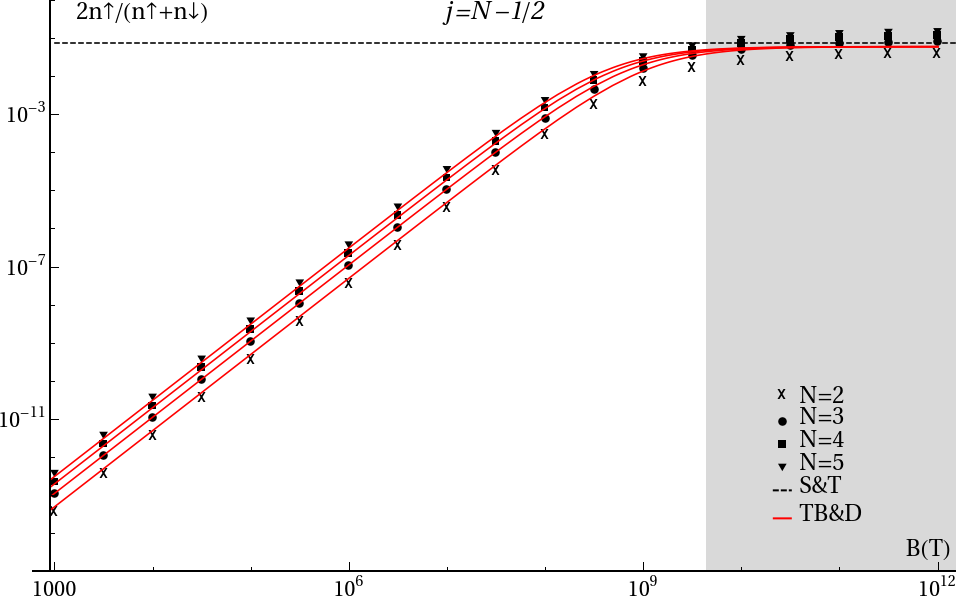}
\caption{The numerically computed deviations from perfect negative spin polarization compared to the S\&T (dashed line) and TB\&D approximations (red curves, $a=0.032$). \label{fig:2}}
\end{figure}

At low field strengths we find a good agreement with the TB\&D approximation for $a=0.032$, where the $N=2$ level spin polarizes a bit better than TB\&D predict, as is shown in FIG.~\ref{fig:2}. For the TB\&D equilibrium spin polarization to converge to the S\&T equilibrium spin polarization, one needs $a=(1-8\sqrt3/15)/2\approx 0.038$, which is close to the values we found by fitting the numerically computed equilibrium spin polarizations for fields up to $10^7$ T, that is, where the S\&T and TB\&D approximations strongly disagree. The value we found is lower than the one for which TB\&D converges to S\&T. This is mostly due to the $N=2$ level having a higher spin polarization than the TB\&D approximation predicts which reduces the fitted value for $a$.

\begin{figure}
\includegraphics[width=\columnwidth]{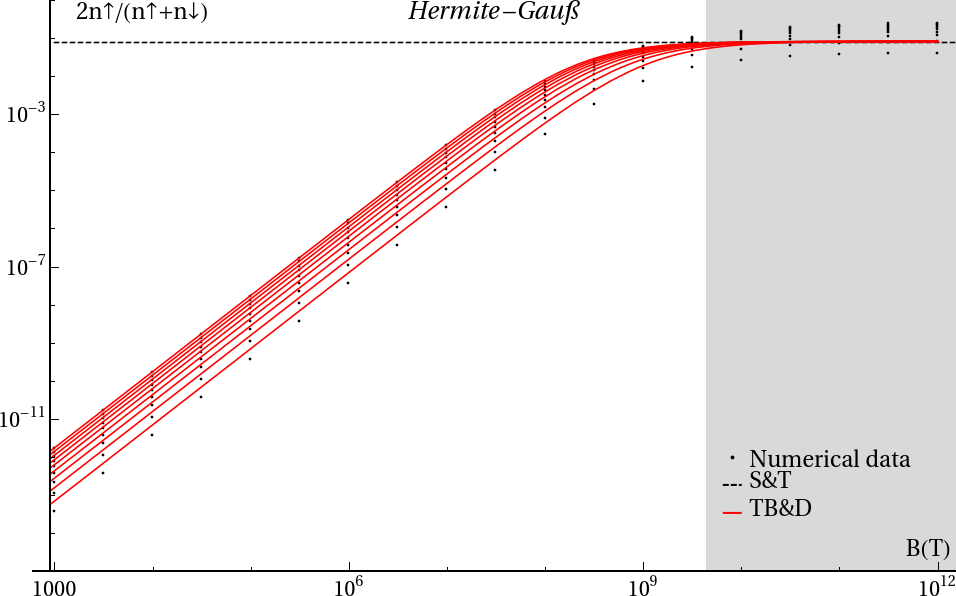}
\caption{The equilibrium deviations from perfect negative spin polarization computed in the Hermite-Gau\ss{} basis. Only the states $N=2$ to $N=10$ are shown for clarity, although states up to $N=60$ were used for fitting the TB\&D approximation (red curves, $a=0.042$). The  $N=2$ equilibrium spin polarization is clearly better than what TB\&D predict.\label{fig:2HG}}
\end{figure}

\begin{figure}
\includegraphics[width=\columnwidth]{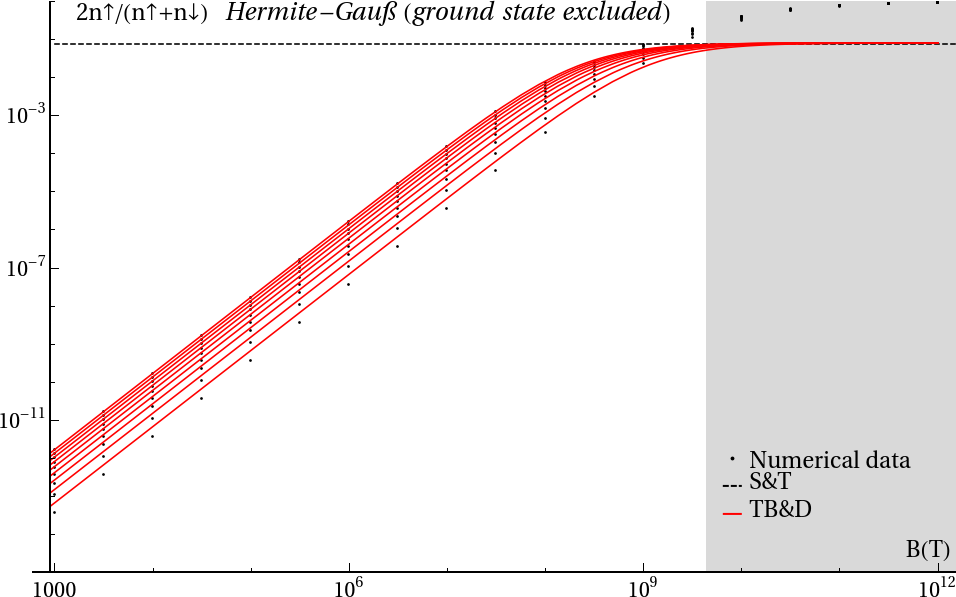}
\caption{The same as FIG.~ \ref{fig:2HG}, but with transitions to the ground state excluded.\label{fig:2HGb}}
\end{figure}

We checked these results against spin flip computations performed in an Hermite-Gau\ss{} basis, where one does not have to sum over angular momenta and thus can compute up to higher quantum numbers efficiently. The results are summarized in FIG.~\ref{fig:2HG}. In the Hermite-Gau\ss{} basis we took states up to $N=60$ and found $a=0.042$. The slightly higher value for $a$ is due to the $N=2$ level contributing less to the fitted value, as the number of states included is much higher. We reran our analysis in the Hermite-Gau\ss{} basis excluding the contributions from transitions to the ground state, which is purely spin polarized against the magnetic field \cite{vanKruiningHayrapetyanGotte17}, to check if a disproportional contribution from ground state transitions is the cause of the discrepancy between the TB\&D approximation and our results. The results of this analysis are summarized in FIG.~\ref{fig:2HGb}. Fitting the data with transitions to the ground state excluded again yielded $a=0.042$ and we found that excluding ground state transitions has little effect on the equilibrium spin polarization at weak magnetic fields. Beyond the critical field we found that transitions to the ground state are almost solely responsible for spin polarizing an electron, as has also been noted in \cite{SokolovZhukovskiNikitina73}. This can be explained because the dominant and spin orbit mixing terms, which have a ratio of $\sqrt{\mathcal E_0+m_e}/\sqrt{\mathcal E_0-m_e}$, become nearly equal for $N\ge1$ in the strong field limit and thus the spin flip rates in both directions should be roughly equal. Only for transitions to the always perfectly spin polarized $N=0$ level does this argument not apply.
%

\begin{figure}
\includegraphics[width=\columnwidth]{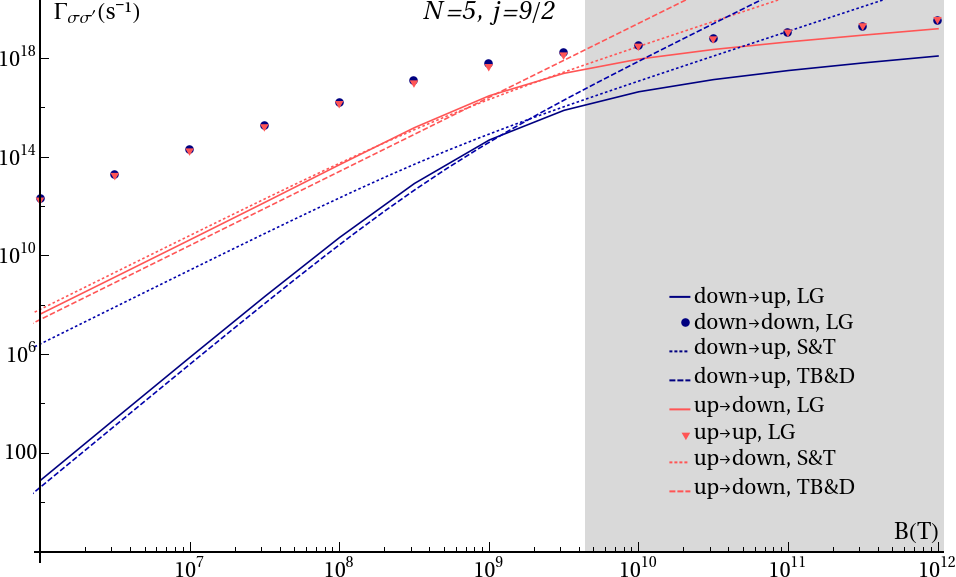}
\caption{Spin flip rates (continuous curves) compared to the S\&T- (dotted curves) and TB\&D approximation (dashed curves) for the initial quantum numbers $N=5$ and $j=9/2$. For low magnetic fields, the transition rate from spin down to spin up is suppressed by many orders of magnitude. Spin preserving decay rates are shown for comparison (dots). Beyond the critical field (gray area), the S\&T- and TB\&D- approximations break down and the spin-flip rates obtained from them diverge from the actual rates.\label{fig:1}}\end{figure}

More pronounced differences between the numerical computations and both approximations show up if one looks at the spin flip rates instead of the equilibrium spin polarization, as is shown in FIG~\ref{fig:1}. Both S\&T and TB\&D overestimate the spin flip rates beyond the critical field.

\begin{figure}
\includegraphics[width=\columnwidth]{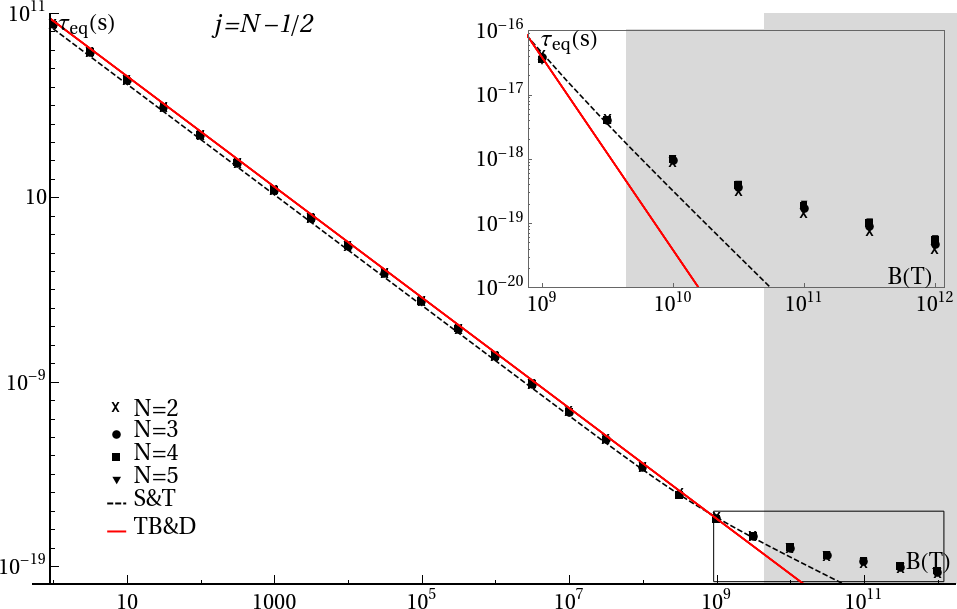}
\caption{Time scale at which spin equilibrium is reached. Interestingly, beyond the critical field (gray area), lower lying states equilibrate faster. Note the diverging S\&T rate beyond the critical field strength, indicating that the S\&T approximation is not applicable in that regime. \textcolor{blue}{}}\label{fig:3}
\end{figure}
\begin{figure}
\includegraphics[width=\columnwidth]{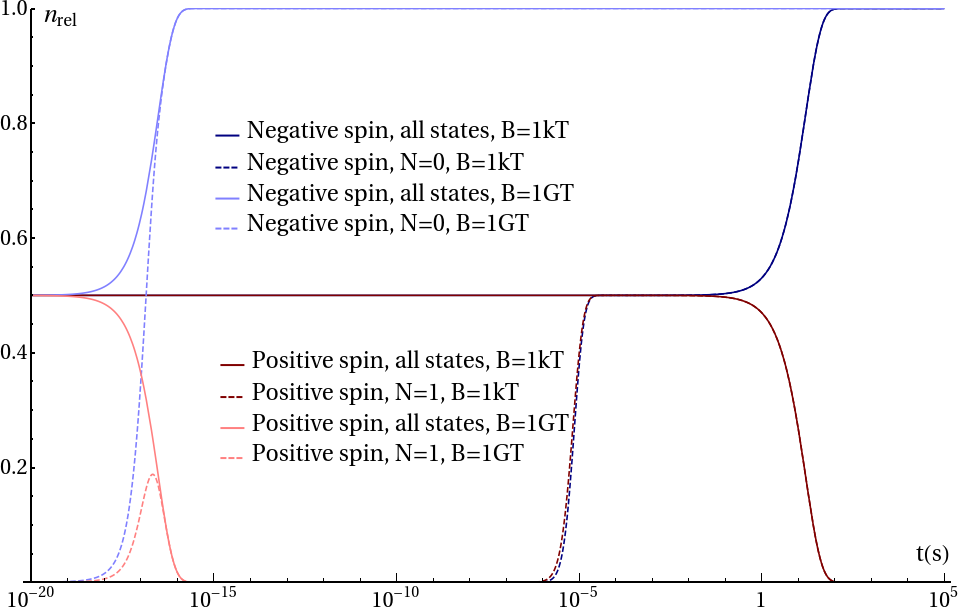}
\caption{Relative occupation of different states starting out from a spin-depolarized $N=5$, $j=9/2$ state for different magnetic field strengths. At low field strength the electron first decays to the lowest state allowed for its spin, before undergoing spin flip at much later times. At high fields the spin flip and spin-preserving decay rates  become comparable and the lowest state for spin up ($N=1$) is less occupied at all times.}\label{fig:4} 
\end{figure}

The overestimation of the spin flip rates shows itself too in the time scale over which an electron reaches its equilibrium spin. In Appendix~\ref{app:teq} we show this time scale to be $\tau_{eq}=1/(\Gamma_{\uparrow\downarrow}+\Gamma_{\downarrow\uparrow})$, the reciprocal of the sum of the spin flip rates. In FIG.~\ref{fig:3} we plot the spin equilibration times for several Landau levels. The equilibration time depends little on the Landau level, but decreases quadratically with $B$ up to the critical field. At one Tesla the spin equilibration time is on the order of a millennium. The actual dynamics are more complicated, with the electron being able to decay to different Landau levels, both via spin-preserving and spin flip decays. At low magnetic fields the time scales of spin preserving and spin flip decays separate, as can be seen in FIG.~\ref{fig:4}. This separation of time scales stems from the spin preserving decays having higher decay rates than the spin flip decays, as one can see in FIG.~\ref{fig:1}. For low magnetic field strengths the electron decays first to an $N=1$ spin up state before undergoing spin flip to an $N=0$ spin down state much later. At high field strengths both processes occur at similar rates and the electron can undergo spin flip before decaying to an $N=1$ spin up state, as is signified by the lower occupation of these states at all times. 

\section{Discussion}
We have shown that in the high field, low excitation limit, existing approximations do not give a good estimate of the electron spin flip rate and its resulting spin polarization. At field strengths beyond the critical field the equilibrium spin polarization is less than the $8\sqrt 3/15$ calculated by S\&T for states excited higher than the second Landau level. 

In our calculations we ignore level shifts due to higher order perturbations, this includes the lifting of the spin-degeneracy due to the electron's anomalous magnetic moment. For the emitted photons we ignore the effects of vacuum birefringence at strong background magnetic fields. Even at the critical field strength the magnitude of these effects is $\sim\alpha/2\pi$ \cite{Constantinescu72a, *Constantinescu72b}. The main effect of the shifting of the electron's energy levels is that the available phase space for the various decay channels change. The decay rates should change proportionally to the available phase space. Therefore we expect the relative changes in the decay rates to be on the order of $\alpha/2\pi$, too. The only qualitatively new feature that can occur due to level shifts is that it allows decays between previously degenerate states, most importantly spin up and spin down with the same $N$. Our method is unable to make predictions about decay rates between such near-degenerate states, but from the small available phase space we expect them to be small even at the critical field strength $O\left((\alpha B/2\pi B_{\text{cr}})^2\right)$. The vacuum birefringence primarily affects the propagation of the emitted photons. The changes it causes in the coupling of the photons to the electron are of second order in $\alpha$. Having considered these effects we believe our calculations are still fairly reliable at the critical field strength.

\section{Outlook}
Our numerical integration over all outgoing photon states assumed a homogeneous density of states, but it can be adapted to inhomogeneous densities as well, making it well suited for cavity QED problems in a strong magnetic field. 


Our results were obtained using transition matrix elements obtained without making any approximation apart from ignoring higher order perturbative effects. To complement our exact approach, for weak magnetic fields the small spatial extent of the electron wave function compared to the wavelength of the radiated photons allows for constructing a simpler approximate model.
\begin{acknowledgments}
We are grateful to Petr Karpov for his help translating articles from Russian.
\end{acknowledgments}
\appendix
\section{explicit expressions for the transition matrix elements}
The transition matrix elements can be computed explicitly by substituting the wave functions for the electron and photon into the expression $\mathcal M=\int\bar\Psi_f\cancel A^*\Psi_idV$. Using cylindrical coordinates, the angular integration gives a factor of $2\pi$ and fixes the angular momentum of the photon. The radial integration can be be performed using the rescaled coordinate $\tilde r=\sqrt{B e/2}r$ and a rescaled transverse wave number $\tilde\kappa=\kappa/\sqrt{2B|e|}$. Taking $Lg_n^l(\tilde r)$ to be the radial part of $LG_n^l(\tilde r, \phi)$, it is convenient to define a shorthand for the radial integrals
\begin{equation}
I_{nn'}^{ll'}(\tilde\kappa)=\frac{4\pi}{B|e|}\int_0^\infty Lg_n^l(\tilde r)J_{l-l'}(2 \tilde \kappa\tilde r)Lg_{n'}^{l'}(\tilde r)\tilde rd\tilde r.
\end{equation}
These integrals can be evaluated to be series of Laguerre-Gau\ss{} functions in $\tilde\kappa$-space \cite{GR}:
\begin{widetext}\begin{multline}
I_{nn}^{ll'}(\tilde\kappa)=\frac{4\pi}{B|e|}e^{-\tilde\kappa^2}\frac{\tilde\kappa^{|l-l'|}}2(-1)^{\frac 12(-|l-l'|-(l-l'))}\sum_{h=0}^{n+n'}\frac{(-1)^h(h+\frac 12(|l|+|l'|-|l-l'|))!}{h!}L_{h+\frac 12(|l|+|l'|-|l-l'|)}^{|l-l'|}(\tilde\kappa^2)\times\\
\sum_{\mu=0}^h\left(\begin{array}{c}h\\ \mu\end{array}\right)
\left(\begin{array}{c}n+|l|\\|l|+\mu\end{array}\right)\left(\begin{array}{c}n'+|l'|\\ |l'|+h-\mu\end{array}\right)
\end{multline}
Using this shorthand and the symbol $\mathcal H$ for the photon helicity ($\mathcal H=1$ for right handed photons and $\mathcal H=-1$ for left handed photons) the matrix elements are
\begin{longtable}{|l|l|l|l|p{14cm}|}\hline
$l$ & $l'$& $\sigma$ & $\sigma'$& $\mathcal M_{\mathcal H}$\\ \hline
$-$ & $-$& $-$& $-$ & $-i\mathcal H\frac \kappa\omega\left(p(\mathcal E'_0+\mathcal E')+p'(\mathcal E_0+\mathcal E)\right)\left(\sqrt{(\mathcal E_0'+m)(\mathcal E_0+m)}I_{nn'}^{ll'}(\tilde \kappa)+\sqrt{(\mathcal E'_0-m)(\mathcal E_0-m)}I_{n-1n'-1}^{l-1l'-1}(\tilde\kappa)\right)$ \newline
$-i\left((\mathcal E'_0+\mathcal E')(\mathcal E_0+\mathcal E)-pp'\right)$\newline
$\left(-\left(1-\mathcal H\frac k\omega\right)\sqrt{(\mathcal E_0'-m)(\mathcal E_0+m)}I_{nn'-1}^{ll'-1}(\tilde\kappa)+\left(1+\mathcal H\frac k\omega\right)\sqrt{(\mathcal E_0'+m)(\mathcal E_0-m)}I_{n-1n'}^{l-1l'}(\tilde\kappa)\right)$
\\ \hline
$-$ & $-$ & $-$ & $+$ & $\mathcal H\frac \kappa\omega\left((\mathcal E_0'+\mathcal E')(\mathcal E_0+\mathcal E)+pp'\right)\left(\sqrt{(\mathcal E_0'+m)(\mathcal E_0-m)}I_{n-1n'}^{l-1l'}(\tilde\kappa)-\sqrt{(\mathcal E_0'-m)(\mathcal E_0+m)}I_{nn'+1}^{ll'+1}(\tilde\kappa)\right)$\newline
$+\left(p(\mathcal E_0'+\mathcal E')-p'(\mathcal E_0+\mathcal E)\right)$\newline
$\left(\left(1-\mathcal H\frac k\omega\right)\sqrt{(\mathcal E'_0+m)(\mathcal E_0+m)}I_{nn'}^{ll'}+\left(1+\mathcal H\frac k\omega\right)\sqrt{(\mathcal E'_0-m)(\mathcal E_0-m)}I_{n-1n'+1}^{l-1l'+1}\right) $
\\ \hline
$-$ & $-$& $+$ & $-$ & $\mathcal H\frac\kappa\omega\left((\mathcal E_0'+\mathcal E')(\mathcal E_0+\mathcal E)+pp'\right)\left(\sqrt{(\mathcal E'_0+m)(\mathcal E_0-m)}I_{n+1n'}^{l+1l'}(\tilde\kappa)-\sqrt{(\mathcal E'_0-m)(\mathcal E_0+m)}I_{nn'-1}^{ll'-1}(\tilde\kappa)\right)$\newline
$+\left(p(\mathcal E_0'+\mathcal E')-p'(\mathcal E_0+\mathcal E)\right)$\newline
$\left(-\left(1-\mathcal H\frac k\omega\right)\sqrt{(\mathcal E'_0-m)(\mathcal E_0-m)}I_{n+1n'-1}^{l+1l'-1}-\left(1+\mathcal H\frac k\omega\right)\sqrt{(\mathcal E'_0+m)(\mathcal E_0+m)}I_{nn'}^{ll'}\right) $
\\ \hline
$-$ & $-$ & $+$ & $+$ & $-i\mathcal H\frac \kappa\omega\left(p(\mathcal E'_0+\mathcal E')+p'(\mathcal E_0+\mathcal E)\right)\left(\sqrt{(\mathcal E_0'+m)(\mathcal E_0+m)}I_{nn'}^{ll'}(\tilde \kappa)+\sqrt{(\mathcal E'_0-m)(\mathcal E_0-m)}I_{n+1n'+1}^{l+1l'+1}(\tilde\kappa)\right)$ \newline
$-i\left((\mathcal E'_0+\mathcal E')(\mathcal E_0+\mathcal E)-pp'\right)$\newline
$\left(-\left(1-\mathcal H\frac k\omega\right)\sqrt{(\mathcal E_0'+m)(\mathcal E_0-m)}I_{n+1n'}^{l+1l'}(\tilde\kappa)+\left(1+\mathcal H\frac k\omega\right)\sqrt{(\mathcal E_0'-m)(\mathcal E_0+m)}I_{nn'+1}^{ll'+1}(\tilde\kappa)\right)$
\\ \hline
$-$ & $+$ & $-$ & $-$ &  $-i\mathcal H\frac \kappa\omega\left(p(\mathcal E'_0+\mathcal E')+p'(\mathcal E_0+\mathcal E)\right)\left(\sqrt{(\mathcal E_0'+m)(\mathcal E_0+m)}I_{nn'}^{ll'}(\tilde \kappa)-\sqrt{(\mathcal E'_0-m)(\mathcal E_0-m)}I_{n-1n'}^{l-1l'-1}(\tilde\kappa)\right)$ \newline
$-i\left((\mathcal E'_0+\mathcal E')(\mathcal E_0+\mathcal E)-pp'\right)$\newline
$\left(\left(1-\mathcal H\frac k\omega\right)\sqrt{(\mathcal E_0'-m)(\mathcal E_0+m)}I_{nn'}^{ll'-1}(\tilde\kappa)+\left(1+\mathcal H\frac k\omega\right)\sqrt{(\mathcal E_0'+m)(\mathcal E_0-m)}I_{n-1n'}^{l-1l'}(\tilde\kappa)\right)$
\\ \hline
$-$ & $+$ & $-$& $+$ & $\mathcal H\frac \kappa\omega\left((\mathcal E_0'+\mathcal E')(\mathcal E_0+\mathcal E)+pp'\right)\left(\sqrt{(\mathcal E_0'+m)(\mathcal E_0-m)}I_{n-1n'}^{l-1l'}(\tilde\kappa)+\sqrt{(\mathcal E_0'-m)(\mathcal E_0+m)}I_{nn'}^{ll'+1}(\tilde\kappa)\right)$\newline
$+\left(p(\mathcal E_0'+\mathcal E')-p'(\mathcal E_0+\mathcal E)\right)$\newline
$\left(\left(1-\mathcal H\frac k\omega\right)\sqrt{(\mathcal E'_0+m)(\mathcal E_0+m)}I_{nn'}^{ll'}-\left(1+\mathcal H\frac k\omega\right)\sqrt{(\mathcal E'_0-m)(\mathcal E_0-m)}I_{n-1n'}^{l-1l'+1}\right) $
\\ \hline
$-$ & $+$ & $+$ &$-$ & $\mathcal H\frac\kappa\omega\left((\mathcal E_0'+\mathcal E')(\mathcal E_0+\mathcal E)+pp'\right)\left(\sqrt{(\mathcal E'_0+m)(\mathcal E_0-m)}I_{n+1n'}^{l+1l'}(\tilde\kappa)+\sqrt{(\mathcal E'_0-m)(\mathcal E_0+m)}I_{nn'}^{ll'-1}(\tilde\kappa)\right)$\newline
$+\left(p(\mathcal E_0'+\mathcal E')-p'(\mathcal E_0+\mathcal E)\right)$\newline
$\left(\left(1-\mathcal H\frac k\omega\right)\sqrt{(\mathcal E'_0-m)(\mathcal E_0-m)}I_{n+1n'}^{l+1l'-1}-\left(1+\mathcal H\frac k\omega\right)\sqrt{(\mathcal E'_0+m)(\mathcal E_0+m)}I_{nn'}^{ll'}\right) $
\\ \hline
$-$ & $+$ & $+$ & $+$ & $-i\mathcal H\frac \kappa\omega\left(p(\mathcal E'_0+\mathcal E')+p'(\mathcal E_0+\mathcal E)\right)\left(\sqrt{(\mathcal E_0'+m)(\mathcal E_0+m)}I_{nn'}^{ll'}(\tilde \kappa)-\sqrt{(\mathcal E'_0-m)(\mathcal E_0-m)}I_{n+1n'}^{l+1l'+1}(\tilde\kappa)\right)$ \newline
$-i\left((\mathcal E'_0+\mathcal E')(\mathcal E_0+\mathcal E)-pp'\right)$\newline
$\left(-\left(1-\mathcal H\frac k\omega\right)\sqrt{(\mathcal E_0'+m)(\mathcal E_0-m)}I_{n+1n'}^{l+1l'}(\tilde\kappa)-\left(1+\mathcal H\frac k\omega\right)\sqrt{(\mathcal E_0'-m)(\mathcal E_0+m)}I_{nn'}^{ll'+1}(\tilde\kappa)\right)$
 \\ \hline
$+$ & $-$ & $-$ &$-$ & $-i\mathcal H\frac \kappa\omega\left(p(\mathcal E'_0+\mathcal E')+p'(\mathcal E_0+\mathcal E)\right)\left(\sqrt{(\mathcal E_0'+m)(\mathcal E_0+m)}I_{nn'}^{ll'}(\tilde \kappa)-\sqrt{(\mathcal E'_0-m)(\mathcal E_0-m)}I_{nn'-1}^{l-1l'-1}(\tilde\kappa)\right)$ \newline
$-i\left((\mathcal E'_0+\mathcal E')(\mathcal E_0+\mathcal E)-pp'\right)$\newline
$\left(-\left(1-\mathcal H\frac k\omega\right)\sqrt{(\mathcal E_0'-m)(\mathcal E_0+m)}I_{nn'-1}^{ll'-1}(\tilde\kappa)-\left(1+\mathcal H\frac k\omega\right)\sqrt{(\mathcal E_0'+m)(\mathcal E_0-m)}I_{nn'}^{l-1l'}(\tilde\kappa)\right)$
\\ \hline
$+$ &$-$ &$-$ & $+$ & $\mathcal H\frac \kappa\omega\left((\mathcal E_0'+\mathcal E')(\mathcal E_0+\mathcal E)+pp'\right)\left(-\sqrt{(\mathcal E_0'+m)(\mathcal E_0-m)}I_{nn'}^{l-1l'}(\tilde\kappa)-\sqrt{(\mathcal E_0'-m)(\mathcal E_0+m)}I_{nn'+1}^{ll'+1}(\tilde\kappa)\right)$\newline
$+\left(p(\mathcal E_0'+\mathcal E')-p'(\mathcal E_0+\mathcal E)\right)$\newline
$\left(\left(1-\mathcal H\frac k\omega\right)\sqrt{(\mathcal E'_0+m)(\mathcal E_0+m)}I_{nn'}^{ll'}-\left(1+\mathcal H\frac k\omega\right)\sqrt{(\mathcal E'_0-m)(\mathcal E_0-m)}I_{nn'+1}^{l-1l'+1}\right) $
\\ \hline
$+$ & $-$& $+$ & $-$ & $\mathcal H\frac\kappa\omega\left((\mathcal E_0'+\mathcal E')(\mathcal E_0+\mathcal E)+pp'\right)\left(-\sqrt{(\mathcal E'_0+m)(\mathcal E_0-m)}I_{nn'}^{l+1l'}(\tilde\kappa)-\sqrt{(\mathcal E'_0-m)(\mathcal E_0+m)}I_{nn'-1}^{ll'-1}(\tilde\kappa)\right)$\newline
$+\left(p(\mathcal E_0'+\mathcal E')-p'(\mathcal E_0+\mathcal E)\right)$\newline
$\left(\left(1-\mathcal H\frac k\omega\right)\sqrt{(\mathcal E'_0-m)(\mathcal E_0-m)}I_{nn'-1}^{l+1l'-1}-\left(1+\mathcal H\frac k\omega\right)\sqrt{(\mathcal E'_0+m)(\mathcal E_0+m)}I_{nn'}^{ll'}\right) $
\\ \hline
$+$ & $-$ & $+$ & $+$ & $-i\mathcal H\frac \kappa\omega\left(p(\mathcal E'_0+\mathcal E')+p'(\mathcal E_0+\mathcal E)\right)\left(\sqrt{(\mathcal E_0'+m)(\mathcal E_0+m)}I_{nn'}^{ll'}(\tilde \kappa)-\sqrt{(\mathcal E'_0-m)(\mathcal E_0-m)}I_{nn'+1}^{l+1l'+1}(\tilde\kappa)\right)$ \newline
$-i\left((\mathcal E'_0+\mathcal E')(\mathcal E_0+\mathcal E)-pp'\right)$\newline
$\left(\left(1-\mathcal H\frac k\omega\right)\sqrt{(\mathcal E_0'+m)(\mathcal E_0-m)}I_{nn'}^{l+1l'}(\tilde\kappa)+\left(1+\mathcal H\frac k\omega\right)\sqrt{(\mathcal E_0'-m)(\mathcal E_0+m)}I_{nn'+1}^{ll'+1}(\tilde\kappa)\right)$
\\ \hline
$+$ & $+$ & $-$& $-$ &  $-i\mathcal H\frac \kappa\omega\left(p(\mathcal E'_0+\mathcal E')+p'(\mathcal E_0+\mathcal E)\right)\left(\sqrt{(\mathcal E_0'+m)(\mathcal E_0+m)}I_{nn'}^{ll'}(\tilde \kappa)+\sqrt{(\mathcal E'_0-m)(\mathcal E_0-m)}I_{nn'}^{l-1l'-1}(\tilde\kappa)\right)$ \newline
$-i\left((\mathcal E'_0+\mathcal E')(\mathcal E_0+\mathcal E)-pp'\right)$\newline
$\left(\left(1-\mathcal H\frac k\omega\right)\sqrt{(\mathcal E_0'-m)(\mathcal E_0+m)}I_{nn'}^{ll'-1}(\tilde\kappa)-\left(1+\mathcal H\frac k\omega\right)\sqrt{(\mathcal E_0'+m)(\mathcal E_0-m)}I_{nn'}^{l-1l'}(\tilde\kappa)\right)$
\\ \hline  
$+$ & $+$ & $-$& $+$ &  $\mathcal H\frac \kappa\omega\left((\mathcal E_0'+\mathcal E')(\mathcal E_0+\mathcal E)+pp'\right)\left(-\sqrt{(\mathcal E_0'+m)(\mathcal E_0-m)}I_{nn'}^{l-1l'}(\tilde\kappa)+\sqrt{(\mathcal E_0'-m)(\mathcal E_0+m)}I_{nn'}^{ll'+1}(\tilde\kappa)\right)$\newline
$+\left(p(\mathcal E_0'+\mathcal E')-p'(\mathcal E_0+\mathcal E)\right)$\newline
$\left(\left(1-\mathcal H\frac k\omega\right)\sqrt{(\mathcal E'_0+m)(\mathcal E_0+m)}I_{nn'}^{ll'}+\left(1+\mathcal H\frac k\omega\right)\sqrt{(\mathcal E'_0-m)(\mathcal E_0-m)}I_{nn'}^{l-1l'+1}\right) $
\\ \hline
$+$ & $+$ & $+$& $-$ & $\mathcal H\frac\kappa\omega\left((\mathcal E_0'+\mathcal E')(\mathcal E_0+\mathcal E)+pp'\right)\left(\sqrt{(\mathcal E'_0-m)(\mathcal E_0+m)}I_{nn'}^{ll'-1}(\tilde\kappa)-\sqrt{(\mathcal E'_0+m)(\mathcal E_0-m)}I_{nn'}^{l+1l'}(\tilde\kappa)\right)$\newline
$+\left(p(\mathcal E_0'+\mathcal E')-p'(\mathcal E_0+\mathcal E)\right)$\newline
$\left(-\left(1-\mathcal H\frac k\omega\right)\sqrt{(\mathcal E'_0-m)(\mathcal E_0-m)}I_{nn'}^{l+1l'-1}-\left(1+\mathcal H\frac k\omega\right)\sqrt{(\mathcal E'_0+m)(\mathcal E_0+m)}I_{nn'}^{ll'}\right) $
\\ \hline
$+$ & $+$ & $+$& $+$ &  $-i\mathcal H\frac \kappa\omega\left(p(\mathcal E'_0+\mathcal E')+p'(\mathcal E_0+\mathcal E)\right)\left(\sqrt{(\mathcal E_0'+m)(\mathcal E_0+m)}I_{nn'}^{ll'}(\tilde \kappa)+\sqrt{(\mathcal E'_0-m)(\mathcal E_0-m)}I_{nn'}^{l+1l'+1}(\tilde\kappa)\right)$ \newline
$-i\left((\mathcal E'_0+\mathcal E')(\mathcal E_0+\mathcal E)-pp'\right)$\newline
$\left(\left(1-\mathcal H\frac k\omega\right)\sqrt{(\mathcal E_0'+m)(\mathcal E_0-m)}I_{nn'}^{l+1l'}(\tilde\kappa)-\left(1+\mathcal H\frac k\omega\right)\sqrt{(\mathcal E_0'-m)(\mathcal E_0+m)}I_{nn'}^{ll'+1}(\tilde\kappa)\right)$
\\ \hline
\end{longtable}
\end{widetext}
\section{Equilibrium spin and equilibration time}\label{app:teq}
To estimate the time scales over which an electron in a magnetic field approaches its equilibrium spin polarization, we use  a simplified two-population model that captures the essentials and is commonly used to derive equlibrium spin polarisations \cite{ST, LL4}. The only assumption we require for this model is that each time after emitting a photon, the electron will be returned to its initial energy level without undergoing further spin changes. Consider a population of $N$ electrons that undergo random spin flips at the rates $\Gamma_{\uparrow\downarrow}$ from spin up to spin down and $\Gamma_{\downarrow\uparrow}$ from spin down to spin up. We furthermore assume that the times at which different electrons undergo spin flips are uncorrelated as are the  times at which one electron undergoes a series of spin flips. The expected spin occupation fractions of spin up electrons $n_\uparrow=N_{\uparrow}/N$ and spin down electrons $n_\downarrow=N_\downarrow/N$ are then given by two coupled first order differential equations (These equations are the same as eq. (21.41) in \cite{ST}).
\begin{equation}\label{eq:popleqs}
\dot n_\uparrow=\Gamma_{\downarrow\uparrow}n_\downarrow-\Gamma_{\uparrow\downarrow}n_\uparrow,\quad \dot n_\downarrow=\Gamma_{\uparrow\downarrow}n_\uparrow-\Gamma_{\downarrow\uparrow}n_\downarrow.
\end{equation}
Obviously the total number of electrons is conserved, $\dot N=N(\dot n_\uparrow+\dot n_\downarrow)=0$. By requiring that the right hand sides of eqs.~(\ref{eq:popleqs}) are zero one can find the equilibrium values of the spin occupation fractions and the spin polarisation
\begin{equation}
n_{\uparrow}=\frac{\Gamma_{\downarrow\uparrow}}{\Gamma_{\uparrow\downarrow}+\Gamma_{\downarrow\uparrow}},\quad n_{\downarrow}=\frac{\Gamma_{\uparrow\downarrow}}{\Gamma_{\uparrow\downarrow}+\Gamma_{\downarrow\uparrow}},\quad
n_{\downarrow}-n_{\uparrow}=\frac{\Gamma_{\uparrow\downarrow}-\Gamma_{\downarrow\uparrow}}{\Gamma_{\uparrow\downarrow}+\Gamma_{\downarrow\uparrow}}
\end{equation}
To find the rate at which the spin polarisation approaches its equilibrium value, we introduce the quantity $\delta n_{eq}$ which is zero when the spin polarisation achieves its equilibrium and is defined as
\begin{multline}
\delta n_{eq}=\frac{\Gamma_{\downarrow\uparrow}n_\downarrow-\Gamma_{\uparrow\downarrow}n_\uparrow}{\Gamma_{\downarrow\uparrow}+\Gamma_{\uparrow\downarrow}}\Longrightarrow \delta\dot n_{eq}=-(\Gamma_{\downarrow\uparrow}+\Gamma_{\uparrow\downarrow})\delta n_{eq}\\
\Longrightarrow \delta n_{eq}\propto e^{-(\Gamma_{\downarrow\uparrow}+\Gamma_{\uparrow\downarrow})t}.
\end{multline}
One can write $\delta n_{eq}$ in the form $n_ 0 e^{-t/\tau_{eq}}$ with $\tau_{eq}$ the equilibration time which is in our case is $1/(\Gamma_{\downarrow\uparrow}+\Gamma_{\uparrow\downarrow})$. 
\bibliography{../../literatuurlijst,../../boeken,Bibliography}
\end{document}